# Constraints on Elliptical Galaxy Formation from the Properties of Their Globular Cluster Systems

Stephen E. Zepf†

*University of California, Berkeley*

**Abstract:** Globular clusters are valuable fossil records of the formation and evolution of their host galaxies. The color distribution of elliptical galaxy globular cluster systems indicates an episodic formation history, consistent with a merger origin for these galaxies. Spectroscopic studies of globular cluster systems are a promising route to further constrain models of elliptical galaxy formation. They also provide a useful probe of the mass distribution of ellipticals at large radii.

## 1. Introduction

The formation mechanism of elliptical galaxies remains an open question. Similarly, the mass distribution of elliptical galaxies is poorly constrained outside of their central regions. These issues are clearly critical to understanding galaxy formation and dark matter, but there is little observational evidence which distinguishes between competing models. One of the primary reasons these questions remain unanswered is that they are difficult to address through studies of the integrated light of elliptical galaxies. The determination of mass-to-light ratios through spectroscopy of the integrated light is limited to within one or two effective radii because of the faint surface brightness of ellipticals outside of their central regions (e.g. Carollo et al. 1995, Bertin et al. 1994). Moreover, studies of stellar populations in integrated light reveal only a luminosity weighted age and metallicity of a stellar population and are unable to distinguish between episodic and monolithic formation histories.

Studies of the globular cluster systems (GCSs) of elliptical galaxies are ideal for addressing the questions of their mass distribution at large radii and the formation history of elliptical galaxies. As bound collections of $10^6$ coeval stars, globular clusters provide a readily observable fossil record of the chemical composition and galactic dynamics at the time of their formation. This is particularly useful for differentiating models, since globular clusters are observed to form efficiently in gas-rich mergers (e.g. Schweizer 1997 and references therein). Moreover, because globular clusters are found at large galactocentric radii, their velocities provide constraints on the mass-to-light ratios of early-type galaxies at these large radii.

---

† Hubble Fellow

## 2. The Formation History from the GCS Color Distribution

In a simple merger scenario for the formation of elliptical galaxies, two populations of globular clusters are expected (Ashman & Zepf 1992). One population is the spatially extended, old metal-poor population associated with the halos of the progenitor spirals, and the second is more spatially concentrated, younger, and more metal-rich population formed in the merger. Because the mergers that form most ellipticals occur at moderate or high redshifts, broad-band colors will primarily reflect metallicity. The metal-rich population formed in the merger should therefore be redder than the metal-poor population of the progenitor spirals (Zepf & Ashman 1993, Zepf, Ashman, & Geisler 1995).

A number of observational programs in the past few years have established that elliptical galaxy GCSs have color distributions with two or more peaks, in agreement with the predictions of the merger models. We refer the reader to the CUP book by Ashman & Zepf (1997) for a full discussion of this subject. Below we plot two of the best-studied cases.

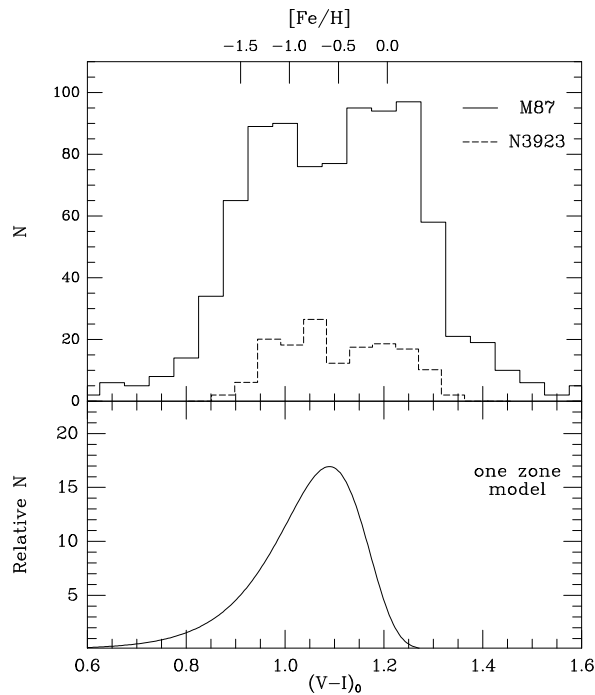

**Figure 1.** The top panel is a plot of the GCS color distributions of M87 (Whitmore et al. 1995) and NGC 3923 (Zepf et al. 1995). The bottom panel is the metallicity distribution expected in a stanard one zone model of chemical evolution, similar to that found in monolithic collapse models for elliptical galaxies (e.g. Arimoto & Yoshii 1987). The GCS color distributions clearly favor episodic formation over single collapse models.

## 3.  Spectroscopic Studies of GCSs

Spectroscopic studies of extragalactic GCSs are beginning to experience a revival, driven both by growing scientific interest and technological advances. Much of the interest in spectroscopy derives from the utility of the radial velocities of clusters as dynamical tracers in elliptical galaxies. Globular clusters are particularly useful in this regard because a significant fraction of them are at large galactocentric radii where it is difficult to obtain other dynamical data.

Determining the metallicities of extragalactic globular clusters is the second motivation for spectroscopy of these objects. The strength of the absorption lines allows a straightforward estimate of the average metallicity. Abundance ratios can also be estimated by comparing absorption lines due to different elements. Although estimating elemental abundance ratios in this way requires data of higher signal-to-noise data than currently typical of spectra of extragalactic globular clusters, it may be possible to use this approach on co-added spectra of a number of clusters. Finally, a comparison of the metallicity estimated from colors and/or spectroscopic line indices with kinematics from velocities can constrain scenarios for the origins of the different globular cluster populations.

Our group is currently analyzing spectra of globular clusters in NGC 4594, NGC 4472, and NGC 3115 obtained primarily at the WHT, and also at the CFHT and KPNO. Our analysis is most complete for the NGC 4594 GCS (Bridges et al. 1996). Using velocities for 34 clusters, we find that $(M/L)_V = 16$ within $5.5'$ (14 kpc). The 90% confidence interval ranges from 11 to 21.5, and assuming completely circular orbits reduces $M/L$ by a factor of 1.5. These can be compared to the results of Kormendy & Westphal (1989), who found $(M/L)_V \simeq 4$ at small radii, with evidence for a gentle rise to about $(M/L)_V \simeq 8$ at $3'$, where all M/L values have been normalized to a distance for NGC 4594 of 8.6 Mpc. Thus, our globular cluster velocities provide evidence for a $M/L$ ratio which rises with radius.

Work is in progress on our spectra of globular clusters around NGC 4472. Our preliminary analysis indicates that we have accurate velocities for more than 40 globular clusters, and that these show no signs of a decline in the velocity dispersion out to a radius of about $5'$ (25 kpc) (Sharples et al. 1997). This would appear to support and extend the result of Mould et al. (1990) who found tentative evidence for a dark halo from the velocities of 26 clusters in the NGC 4472 GCS. We will also make a detailed comparison of kinematics with galaxy color and absorption-line strength, in order to look for any kinematic differences between the red and blue populations clearly dilineated in photometric studies of the NGC 4472 GCS (Geisler et al. 1996, Zepf & Ashman 1993).

Although the numbers are very small, it may be interesting to note that the few early-type galaxies for which there are velocities for both globular clusters and planetary nebulae give, $(M/L)_{\rm GCS} \simeq (M/L)_{\rm PNe}$, but $(v_{max}/\sigma)_{\rm GCS} > (v_{max}/\sigma)_{\rm PNe}$. These are based on NGC 5128 (Hui et al. 1994, Harris, Harris, & Hesser 1988, Sharples 1988), NGC 1399 (Arnaboldi et al. 1994, Grillmair et al. 1994) and NGC 4594 (Freeman et al. 1996, Bridges et al. 1996). Given the small numbers involved, it is probably best to avoid speculation and note that in the near future, the number of galaxies for which this comparison is possible should grow rapidly.

I thank my collaborators on the various projects described above, including Ray Sharples, Dave Hanes, and Terry Bridges who have played major roles in the spectroscopic programs, and Keith Ashman who has been my collaborator since we began working on globular clusters. My research is supported by NASA through grant number HF-1055.01-93A awarded by the Space Telescope Science Institute, which is operated by the Association of Universities for Research in Astronomy, Inc., for NASA under contract NAS5-26555. I also acknowledge the support of an AAS travel grant.